\documentclass[12pt,twoside]{article}
\usepackage{correl}

\def\TheTitle{Introduction to Mean-Field Theory of Spin Glass Models}
\def\TheHeading{Mean-Field Spin-Glass Models}
\def\TheAuthor{V\'aclav Jani\v s}
\def\TheInstitute{Institute of Physics, Academy of Sciences of the Czech
  Republic,  Prague}
\def\TheAddress{}
\def\TheChapter{5}                       

\begin{document}
\MakeTitle           
\section{Introduction}

\subsection{Electronic background of spin-glass interactions}

Electrons are responsible for most of the low-temperature electro-magnetic phenomena in solids. That is why there is a vast effort to apprehend reliably the behavior of systems with many electrons in crystals. Electrons possess charge and spin, and since their mass is small, also non-negligible kinetic degrees of freedom. Moreover, at low temperatures the Fermi statistics and the Pauli principle add to complexity of many-electron systems. Combination of the Coulomb repulsion, kinetic energy and Fermi statistics leads to a large scale of collective quantum phenomena in which electrons are either major agents or act indirectly as mediators. The latter is the case of metallic spin glasses.         

Spin glasses are magnetic systems in which the interaction between the well formed and immobile magnetic moments is frustrated \index{exchange interaction!random frustrated} and the forces from different sources are in conflict with each other due to a frozen structural disorder \index{disorder!frozen}. Hence, no conventional long-range order establishes at low temperatures. Nevertheless, these systems exhibit a freezing or ordering temperature signaling emergence of a new ``glassy '' phase.  The classical examples of spin glasses are noble metals (Au, Ag, Cu, Pt) weakly diluted with transition metal ions, mostly Fe or Mn. Although only the interaction between the local magnetic moments of the transition metal ions is relevant, the metal matrix in which they are diluted supplies the mediating particles, conduction electrons. The scattering of the conduction electrons on magnetic impurities leads to an indirect Ruderman, Kittel, Kasuya and Yosida (RKKY) \cite{Ruderman:1954aa,Kasuya:1956aa,Yosida:1957aa} exchange interaction between the spins that strongly oscillates with distance $\mathbf{R}$,        
\begin{equation*}
J(\mathbf{R}) = J_{0}\ \frac{\cos(2k_{F}R + \phi_{0})}{\left( k_{F}R\right)^{3}} \ , \quad R\to \infty\ .
\end{equation*}
Here $J_{0}$ and $\phi_{0}$ are material constants and $k_{F}$ is the Fermi wave number of the host metal. Since the magnetic ions are strongly diluted in the host, the distances between the spins are effectively random leading to a random distribution of the spin exchange with no preference for a homogeneous magnetic order. Positive and negative signs of the exchange are then equally likely and we can model the RKKY interaction by a randomly fluctuating exchange $J(R)= \pm J_{0}/k_{F}R^{3}$. Since the alloys are very dilute, the average distance between the active spins is large and a mean-field approximation is well justified. In this way the first mean-field model of spin glasses was proposed by Edwards and Anderson \cite{Edwards:1975aa}, just a few years after the first experimental evidence of spin-glass behavior was discovered \cite{Cannella:1972aa}.

\subsection{Models of spin glasses}

The spin of the conduction electron of the noble metals serves only to mediate the interaction between the local spin moments of the transition metal alloys. In modeling the spin-glass behavior we can resort to purely immobile spins distributed regularly on a lattice with a random spin exchange with no preference for either ferro or antiferromagnetic long-range order. If $H[J,S]$ is a spin Hamiltonian we have to evaluate the free energy of the system in the thermodynamic limit. We assume that the free energy is self-averaging and the system is ergodic, hence     
 \begin{equation}\label{eq:AFE-general}
- \beta F(T) = \lim_{N\to\infty}\ln \text{Tr}_{S} \exp\left\{ -\beta H[J,S]\right\} = \left\langle \ln \text{Tr}_{S} \exp\left\{ -\beta H[J,S]\right\} \right\rangle_{J} \ .
\end{equation}
We will later analyze whether any of the fundamental assumptions of the standard statistical mechanics is broken in the mean-field theory of spin glasses. Although Heisenberg spins are physically most interesting, the glassy behavior in random frustrated spin systems becomes so complex that the full understanding of the new phenomena having origin in the glassy phase demands either further simplifications or alternative spin models to study. We will deal with three spin-glass models that show different scenarios of the paramagnetic to spin-glass transition.

\subsubsection{Heisenberg and Ising models}

The simplest lattice spin system consists of spins with the lowest value $\hbar/2$. We set $\hbar/2 = 1, k_{B} = 1$ to simplify the resulting formulas.  Since the transition temperature is relatively high, the spins can be treated classically. The Hamiltonian of the Heisenberg spin model in a homogeneous external magnetic field $\mathbf{h}$ on a regular lattice is
\begin{equation}\label{eq:H-Heinseberg}
 H [J,\mathbf{S}]= -\sum_{i<j}^{N} J_{ij} \mathbf{S}_i\cdot \mathbf{S}_j  - \sum_{i}\mathbf{h}\cdot\mathbf{S}_{i} 
 \end{equation}
with a normalization condition $\mathbf{S}_{i}\cdot\mathbf{S}_{i}=1$.  Although the spin rotational degrees of freedom influence the way the system can go from the paramagnetic to the spin-glass phase \cite{Gabay:1981aa,Cragg:1982aa}.  It is actually the Ising model that has been mostly studied in theories of spin glasses. Only the projection $S^{z}$ of the spin vector to the easy axis, determined by the external magnetic field, are significant in the Ising model. \index{mean field model!Ising glass} We use the Ising model as the generic case for the demonstration of properties of the mean-field solutions of spin glasses. The mean-field limit of the Ising spin glass is called Sherrington-Kirkpatrick (SK) model.  \index{Sherrington-Kirkpatrick!model}

\subsubsection{Potts model}

The $p$-state Potts model is a generalization of the Ising model to $p> 2$ spin components.  The original formulation of Potts~\cite{Potts:1952aa} with
 Hamiltonian $H_{p}=-\sum_{i<j}J_{ij}\delta_{n_{i},n_{j}}$ where
$n_{i}=0,\ldots, p-1$ is an admissible value of spin projections of the $p$-state
model on the lattice site $\mathbf{R}_{i}$, is unsuitable for
practical calculations. The Potts Hamiltonian can, however, be represented via
interacting spins~\cite{Wu:1982aa} 
 \begin{align}\label{eq:H-Potts}
H_{P}\left[J,\mathbf{S}\right] &=-\frac{1}{2}\sum_{i,j}^{N}J_{ij}\mathbf{S}_{i}\cdot \mathbf{S}_{j}  - \sum_{i}\mathbf{h}\cdot \mathbf{S}_{i}\ ,
\end{align} 
where $\mathbf{S}_{i} = \{s_{i}^{1},\ldots s_{i}^{p-1} \}$ are Potts
vector variables taking values from a set of state vectors
$\{\mathbf{e}_{A}\}_{A=1}^{p}$. Functions on vectors $\mathbf{e}_{A}$
are in equilibrium fully defined through their scalar product
\begin{align*}
\sum_{A=1}^{p}e^{\alpha}_{A}=0\ , &\quad  
\sum_{A=1}^{p}e^{\alpha}_{A}e^{\beta}_{A}=p\ \delta^{\alpha \beta} \ , & e^{\alpha}_{A}e^{\alpha}_{B}=p\ \delta_{A B}-1
\end{align*}
for  $\alpha\in\{1,...,p-1\}$. We use the Einstein summation convention for
repeating Greek indices of the vector components indicating a scalar
product of the Potts vectors. \index{mean field model!Potts glass} Using these properties we can construct an explicit representation of the Potts spin vectors
\begin{align*}
e^{\alpha}_{A}=\left\{
\begin{array}{ll}
   0 & A<\alpha \\
   \sqrt{\frac{p(p-\alpha)}{p+1-\alpha}}  & A=\alpha \\
   \frac{1}{\alpha-p}\sqrt{\frac{p(p-\alpha)}{p+1-\alpha}} &
   A>\alpha\ .
\end{array}
\right. 
\end{align*}
The Potts model with a random spin exchange shows a transition to a glassy phase but the scenario depends on the number of spin components $p$.

\subsubsection{$p$-spin model} 
 
Potts model is not the only interesting extension of the Ising model. Another generalization is the so-called $p$-spin model. It describes a system of Ising spins where the spin exchange connects a cluster of $p$ spins.    The Hamiltonian of such a model reads  \cite{Gross:1984aa}
 \begin{align}\label{eq:H-p-spin}
H_{p}\left[J,S\right] &=\sum^{N}_{1\le i_{1} < i_{2} < \ldots <i_{p}}J_{i_{1}i_{2}\ldots i_{p}}S_{i_{1}}S_{i_{2}}\ldots S_{i_{p}} \ .
\end{align}
The $p$-spin model is interesting in that we can analytically study the limit $p\to\infty$ for which we know the exact solution being the random-energy model~\cite{Derrida:1980aa,Derrida:1981aa}.  It is equivalent to the one-level replica-symmetry breaking solution from the replica trick \cite{Mezard:1984aa}. There were hopes that one could understand better the genesis of the full mean-field solution of the Ising glass by using the inverse number of the coupled spins $1/p$ as a small parameter starting from the random-energy model. We show later on that $1/p$ expansion does not work, since it does not cure negative entropy at zero temperature for $p<\infty$. \index{mean field model!p-spin glass}

\subsection{Replica trick and the Sherrington-Kirkpatrick mean-field solution}
 
Having the spin Hamiltonian one has to resolve the free energy by evaluating the right-hand side of Eq.~\eqref{eq:AFE-general}. We use the regular lattice with a frustrated random nearest-neighbor interaction to simulate the magnetic properties of the diluted magnetic ions in noble metals. Since the average distance between the ions is large, one assumes a long-range spin exchange in the spin-glass models which leads us naturally to a mean-field theory. The modern understanding of  the mean-field approximation is the mathematical limit $d\to\infty$ of the model on a $d$-dimensional hypercubic lattice. Lattice mean-field theories with long-range interactions are also called models on fully connected graphs, where each node of the graph is connected to any other node.      

The limit to infinite dimensions or the long-range interaction introduces a new large scale. To make the thermodynamic limit meaningful the dependence of the energy on this new large scale must be compensated by rescaling the non-local spin exchange so that the energy remains linearly proportional to volume or the number of lattice sites (spins). Since  the linear contribution from the spin exchange $J_{ij}$ is missing in the spin-glass models  due to frustration that does prefers ferromagnetic nor antiferromagnetic order we have to rescale it as $J = J_{ij}/\sqrt{N}$. To simplify the numerics one usually chooses a Gaussian random distribution  
 \begin{align}\label{eq:MF-Gauss}
 P(J_{ij}) & = \sqrt{\frac{N}{2\pi J^{2}}} \ \exp\left\{- \frac{N J_{ij}^{2}}{2 J^{2}} \right\} \ .
 \end{align} \index{exchange interaction!random frustrated}
Even with the mean-field simplification one faces a hard problem of averaging the logarithm of the partition sum, that is to evaluate the following multiple integral 
 \begin{equation}\label{eq:F-av}
-\beta \left\langle F_{N}\right\rangle_{av} = \int \prod\limits_{i<j}^{N}d[J_{ij}] P[J_{ij}] \ln \int \prod\limits_{i=1}^{N}
d[\mathbf{S}^a_i]\rho[\mathbf{S}^a_i] \exp\left\{ -\beta
H[J,\mathbf{S}]\right\} \ .
\end{equation}
The next simplification is introduced by the application of the replica trick \index{replica!trick} introduced in Ref.~\cite{Edwards:1975aa}. We create $\nu$ copies of the original spin variables and average the replicated system over the random spin exchange. Simply counting the diagrammatic contributions to the free energy where the spin exchange is represented by a bond, one easily finds that each closed loop contributes $\nu$ times. Hence, the free energy can be represented as     
 \begin{align}\label{eq:replica-trick}
\beta F & = - \lim_{\nu \to 0}\left[ \frac 1{\nu }\lim_{N\to\infty}\left( \left\langle Z^\nu_{N}\right\rangle_{av} - 1 \right)\right] \ .
\end{align}
Representation~\eqref{eq:replica-trick} simplifies the averaging over randomness tremendously, in particular when the Gaussian distribution of the random spin exchange from Eq.~\eqref{eq:MF-Gauss} is used. The integration over the spin exchange can be explicitly performed for the replicated partition sum and then, assuming \textit{ergodicity} \index{ergodicity} and the existence of the thermodynamic limit together with validity of linear response, one can get an explicit representation for the density of the free energy of the Ising spin glass with a single order parameter $q = N^{-1}\sum_{i}m_{i}^{2}$ in the glassy phase  
\begin{align}\label{eq:FE-SK}
f(q) = -\frac \beta4(1-q)^2  -\frac 1{\beta}
\int_{-\infty}^{\infty}\mathcal{D}\eta\ln 2\cosh\left[\beta\left(h +
\eta\sqrt{q}\right)\right] 
\end{align}
derived by Sherrington and Kirkpatrick \cite{Sherrington:1975aa}.  \index{Sherrington-Kirkpatrick!solution}

At first sight, there is nothing wrong with the derivation of the Sherrington-Kirkpatrick free energy, hence one would not expect any unphysical behavior. Nevertheless, already the authors themselves found that entropy, calculated from free energy as $S(T) = - \partial F(T)/\partial T$ leads to a negative value at zero temperature, \index{entropy!negative} $S(0) = - \sqrt{\frac 2\pi}\ k_{B} \approx - 0.798k_{B}$. And this is too bad. This unexpected result triggered an avalanche of attempts to resolve the enigma. It was found soon by Monte-Carlo simulations that the entropy of the Sherrington-Kirkpatrick model is non-negative and approaches zero when lowering the temperature \cite{Kirkpatrick:1977aa,Kirkpatrick:1978aa}. Analyzing the stability of the SK solution de Almeida and Thouless  showed that it is actually unstable in the whole spin-glass phase. The stability condition to be satisfied  by the SK solution    
\begin{align}\label{eq:AT-condition}
\Lambda &= 1 - \beta^{2}\left\langle\left(1 - \tanh^{2}\left[\beta\left(h +
\eta\sqrt{q}\right)\right]\right)^{2} \right\rangle_{\eta} \ge 0   
\end{align}
is broken everywhere below the transition temperature along the de Almeida-Thouless (AT) line \index{de Almeida-Thouless!line} in arbitrary magnetic field \cite{deAlmeida:1978aa}. Initially it was suggested that the replica trick and averaging over the spin exchange prior to averaging the thermal fluctuations is responsible for the instability \cite{vanHemmen:1979aa}. The theory of Thouless, Anderson and Palmer, where thermal averaging was performed for a fixed configuration of the spin exchanges, did not, however, resolve the problem \cite{Thouless:1977aa}. When averaging over the random exchange is applied  in their theory by assuming ergodicity,  \index{ergodicity}one ends up with the SK solution \cite{Janis:2006aa}. 
After a number of unsuccessful or only partially successful attempts it was Giorgio Parisi  \index{Parisi!Giorgio} who proposed a scheme of replica-symmetry breaking that would lead to a (marginally) stable and thermodynamically consistent equilibrium state \cite{Parisi:1979aa,Parisi:1979ab,Parisi:1980ac,Parisi:1980ab,Parisi:1980ae,Parisi:1980ag}. Indeed, two decades later it was rigorously proven that the Parisi construction of the replica symmetry breaking leads to the exact free energy of the Sherrington-Kirkpatrick model \cite{Guerra:2003aa,Talagrand:2006aa}.       

There is vast literature documenting the way to the full mean-field solution of the spin glass models and we refer the reader to review articles \cite{Binder:1986aa,Zamponi:2010aa} or books  \cite{Mezard:1987aa,Fischer:1991aa,Dotsenko:2001aa,Nishimori:2001aa,DeDominicis:2006aa} for details. Here we use an offbeat road to the full solution of the mean-field spin-glass models by trying to identify the thermodynamic origin for the  failure of the SK mean-field solution \cite{Janis:2005aa,Janis:2008aa,Janis:2013aa,Janis:2015aa}.

\section{Fundamental concepts of the full mean-field theory of spin glasses}

We now know that the replica trick and the way we use it is not the cause of the instability of the SK mean-field solution. One has to find a physical argument or a hole in the derivation of the SK free energy. Although the SK solution is unstable within the whole glassy phase, the major problem of this simple mean-field theory is negative entropy at very low temperatures. This constitutes a severe intolerable problem. One has to find the reason for this behavior.  In deriving the mean-field solution we assumed that the fundamental principles of statistical mechanics are valid, the thermodynamic limit exists and different statistical ensembles are equivalent. Only if this is true we can equal the entropy calculated from the free energy of the canonical ensemble with the entropy from the microcanonical one. The entropy in the latter ensemble is positive from definition and hence the canonical and microcanonical ensembles in the Sherrington-Kirkpatrick solution are not equivalent and do not lead to the same physical results. The assumptions for the existence of the thermodynamic limit must then be revisited.

\subsection{Ergodicity, thermodynamic limit and thermodynamic homogeneity}

The very fundamental basis on which statistical mechanics is built is the ergodic hypothesis. It is the means by which the long-time development of a microscopic state is related to statistical averaging over the allowed states in the phase space. The Birkhoff ergodic theorem \index{ergodic!theorem} asserts that for ergodic systems \cite{Birkhoff:1931aa} 
\begin{align*}
\left\langle f\ \right\rangle_{T} \equiv \lim_{T\to \infty} \frac 1T \int _{t_{0}}^{t_{0}+ T} f\ (X(t)) dt = \frac{1}{\Sigma_{E}} \int_{S_{E}} f\ (X) d S_{E} &  \equiv \left\langle f\ \right\rangle_{S} \ , 
\end{align*}
holds. Here $T$ is a macroscopic time scale, $X(t)$ is the classical trajectory of a microscopic state, $S_{E}$ is the a constant energy subspace of the phase space and $\Sigma_{E}$ its volume. It means that the classical trajectory of the microscopic state covers almost everywhere the allowed phase space constrained by external macroscopic conditions. It is, however, highly nontrivial to determine which are the allowed states indeed with only homogeneous macroscopic parameters. This is actually the most difficult task in constructing the proper phase space in statistical models. That is, to find out which points of the phase space are infinitesimally close to the trajectory of the many-body system extended to infinite times. Since we never solve the equation of motion of the statistical system, we have only static means to check the validity of ergodicity. \index{ergodicity} We do it by testing the validity of consequences of the ergodic hypothesis and its impact on the behavior of the equilibrium state. The most important consequence of ergodicity of statistical systems is the existence of the thermodynamic limit. \index{thermodynamic!limit}    
 
The trajectory of the many-body system covers almost the whole allowed phase space. It means that the space covered by such trajectory does not depend on the initial state in non-chaotic systems. In ergodic systems then the thermodynamic limit does not depend on the specific form of the volume in which the macroscopic state is confined as well as on its surrounding environment. The ergodic macroscopic systems can either be isolated or embedded in a thermal bath. The thermodynamic equilibrium, the equilibrium state in the thermodynamic limit, is the same with vanishing relative statistical fluctuations. The thermodynamic equilibrium can then be reached by limiting any partial volume of the whole to infinity. The ergodic equilibrium state is homogeneous in the thermodynamic limit.        

Thermodynamic homogeneity \index{thermodynamic!homogeneity} is usually expressed via Euler's lemma~\cite{Reichl:1980aa}
\begin{equation*}
\alpha\ F(T,V,N,\ldots,X_i,\ldots) =  F(T, \alpha V,
\alpha N,\ldots,\alpha X_i,\ldots)
\end{equation*}
telling us that the thermodynamic potential, free energy $F$ in this case, is an extensive variable and is a first-order homogeneous function  of all its extensive variables, volume $V$ number of particles $N$, and the model dependent other extensive variables $X_{i}$. As a consequence of Euler's lemma we obtain that thermodynamic equilibrium is attained as a one-parameter scaling limit where we have only one independent large scale, extensive variable, be it either volume or number of particles, and the other extensive variables enter the thermodynamic potentials as volume or particle densities insensitive to fluctuations of the scaling variable. 

Although the foundations of statistical mechanics are based on ergodicity, lack of ergodicity is widespread in physical phenomena~\cite{Palmer:1982aa}. Typical examples of ergodicity breaking \index{ergodicity!breaking} are phase transitions with a symmetry breaking in the underlying Hamiltonian. Broken ergodicity is sometimes used as a generalization of spontaneous symmetry breaking~\cite{Bantilan:1981aa}.  Although broken global symmetry is always accompanied by broken ergodicity, the converse does not hold. Ergodicity is broken in the mean-field spin glass models without any symmetry of the Hamiltonian being simultaneously broken.  

Broken ergodicity represents an obstruction in the application of fundamental thermodynamic laws. It hence must be recovered. When ergodicity is broken in a phase transition breaking a symmetry of the Hamiltonian, one introduces a symmetry breaking field into the Hamiltonian, being the Legendre conjugate to the extensive variable that is not conserved in the broken symmetry transformation in the low-temperature phase. The symmetry-breaking field allows one to circumvent the critical point of the symmetry-breaking phase transition and simultaneously restores ergodicity. Systems with no broken symmetry of the Hamiltonian at the phase transition do not offer natural external symmetry-breaking fields. Other techniques must be employed to find the proper portion of the phase space covered by the trajectory of the microscopic state in the low-temperature phase and to restore ergodicity.


\subsection{Real replicas and phase-space scalings}

Thermodynamic homogeneity allows us to use scaling of the original phase space. Thermodynamic quantities remain unchanged if we arbitrarily rescale the phase space and then divide the resulting thermodynamic potential by the chosen scaling (geometric) factor. We can do that by scaling the energy $E$ of the equilibrium state. If we use a scaling factor $\nu$, that can be an arbitrary positive number,  then the following identities hold for entropy $S(E)$ of the microcanonical and free energy $F(T)$ of the canonical ensemble  with energy $E$ and temperature $T$, respectively  
\begin{subequations}
\begin{align}\label{eq:Homogeneity}
S(E) &= k_B \ln \Gamma(E) =  \frac {k_B}\nu \ln \Gamma(E)^\nu =\ \frac{ k_B}
\nu \ln \Gamma(\nu E) \ , \\ F(T) &= -\ \frac {k_BT}\nu  
\ln\left[\text{Tr}\ e^{-\beta H}\right]^\nu = -\ \frac {k_BT}\nu  
\ln\left[\text{Tr}\ e^{-\beta \nu H}\right]\ ,
\end{align}
\end{subequations}
where we denoted  by $\Gamma(E)$ the phase-space volume of the isolated system with energy~$E$.

The scaling of the phase space with an integer scaling factor $\nu$ can be simulated by replicating $\nu$-times the extensive variables. That is, we use instead of a single phase space $\nu$ replicas of the original space.  The reason to introduce replicas  \index{replica!real}of the original variables is to extend the space of the available states in the search for the allowed space in equilibrium. The replicas are independent when introduced. We use the replicated variables to study the stability of the original system with respect to fluctuations in the surrounding thermal bath. For this purpose we break the independence of the replica variables by switching on a (homogeneous) infinitesimal interaction between the replicas that we denote $\mu^{ab}$. We then add a small interacting part $\Delta H(\mu)=  \sum_{i} \sum_{a < b}^{\nu}  \mu^{ab} X_i^a X_i^b$ to the replicated  Hamiltonian with dynamical extensive variables $X_{i}$. The original system is then stable with respect to fluctuations in the bath, represented by the interaction between the replicated variables, if the linear response to perturbation $\mu$ is not broken.  If the linear response holds then the perturbed free energy per replica relaxes, after switching perturbation $\mu$ off, to the original one in the thermodynamic limit 
\begin{align}\label{eq:avFE}
  - \beta F_\nu(\mu)  &= \ \frac 1\nu
 \ln\mathrm{Tr}_{\nu}\exp\left\{-\beta\sum_{a=1}^{\nu} H^a -\beta
      \Delta H(\mu) \right\}
\xrightarrow[\mu \to 0]{} \ln\mathrm{Tr}\exp\left\{-\beta H\right\} \ , 
\end{align}
where $\mathrm{Tr}_{\nu}$ refers to the trace in the $\nu$-times replicated phase space. \index{replica!phase space} If the linear response to the inter-replica interaction is broken, the thermodynamic limit of the original system is not uniquely defined and depends on properties of the thermal bath represented by the replicated variables. If there are no apparent physical fields breaking the symmetry of the Hamiltonian, the phase-space scaling represented by replicas of the dynamical variables introduces shadow or auxiliary symmetry-breaking fields, \index{fields!symmetry-breaking} inter-replica interactions \index{replica!interactions} $\mu^{ab}>0$. They induce new order parameters in the response of the system to these fields that need not vanish in the low-temperature phase, when the linear response breaks down. The real replicas offer a way to disclose a degeneracy when the thermodynamic limit is not uniquely defined by a single extensive scale and densities of the other extensive variables. The inter-replica interactions are not measurable and hence to restore the physical situation we have to switch off these fields at the end. If the system is thermodynamically homogeneous we must fulfill the following identity
\begin{equation}\label{eq:av-homogeneity}
\lim_{\mu\to0} \frac {d F_\nu (\mu)}{d\nu}  \equiv 0
\end{equation}
for arbitrary $\nu$. This quantification of the global thermodynamic homogeneity, \index{thermodynamic!homogeneity!global} thermodynamic independence of the scaling parameter $\nu$, will lead us to the construction of a stable solution of mean-field spin glass models. To use equation~\eqref{eq:av-homogeneity} in the replica approach we will need to  analytically continue the replica-dependent free energy to arbitrary positive scaling factors $\nu \in \mathbb R^{+}$. Specific assumptions on the symmetry of matrix $\mu^{ab}$ will have to be introduced. It is evident from Eq.~\eqref{eq:avFE} that the linear response to inter-replica interactions can be broken only if the replicas are mixed in the $\nu$-times replicated free energy $F_{\nu}$.

\subsubsection{Replicated Sherrington-Kirkpatrick model}

We apply real replicas to test thermodynamic homogeneity of the Sherrington-Kirkpatrick model. We replicate the Ising Hamiltonian $\nu$ times
\begin{equation*}
[H]_\nu = \sum\limits_{a=1}^{\nu}H^{a}=
\sum\limits_{\alpha=1}^{\nu}\sum_{<ij>}J_{ij}S_i^a S_j^a
\end{equation*}
and add a small replica-mixing perturbation $\Delta H(\mu)= \frac 12 \sum_{a\neq b} \sum_i \mu^{ab} S_i^a S_i^b$. We assume that ergodicity and linear response hold in the replicated phase space. The averaging over the long-range spin exchange leads to mixing of the replicated spins and after performing averaging over the random spin exchange we obtain in the limit $\mu\to 0$ an extended Sherrington-Kirkpatrick  free-energy density with the Sherrington-Kirkpatrick order parameter $q$ and new off-diagonal parameters $\chi^{ab}$, \cite{Janis:2005aa} \index{Sherrington-Kirkpatrick!order parameter} \index{Sherrington-Kirkpatrick!model!replicated}
\begin{multline}\label{eq:FE-averaged-finite}
f_\nu = \frac{\beta J^2}{4} \left[\frac 1\nu\sum_{a\neq b}^\nu
\left\{\left(\chi^{ab}\right)^2 + 2 q\chi^{ab}\right\}
- (1 - q)^2\right]\\ -\frac
1{\beta\nu}\!\int\limits_{-\infty}^{\infty}\frac{d\eta}{\sqrt{2\pi}} \
e^{-\eta^2/2} \ln \text{Tr}_{\nu} \exp\left\{\beta^2J^2\sum_{a < b}^\nu
\chi^{ab}S^aS^b + \beta \bar{h}\sum_{a=1}^\nu S^a\right\} \ .
\end{multline}
The new parameters in the extended phase space are the response functions conjugate to the inter-replica interaction $\mu^{ab}$ and are the inter-replica susceptibilities \index{replica! overlap susceptibilities} $\chi^{ab} = \langle\langle S^a S^b\rangle_T\rangle_{av} - q$, in a complete analogy to the real magnetic field to which magnetic susceptibility is a linear response. Since we deal with a mean-field model, the susceptibilities are local. Here $\langle f(S)\rangle_{T}$ denotes averaging over thermal fluctuations and $\langle f(S)\rangle_{av}$ over the spin exchange. The SK order parameter is  $q = \langle\langle S^a\rangle_T^2 \rangle_{av}$ and $\bar{h} = h + \eta\sqrt{q}$.  

Replicated spin variables introduced new order parameters that should be determined from stationarity of the free-energy density from Eq.~\eqref{eq:FE-averaged-finite}. There are, however, two problems with this free energy. First, it is not a closed expression that would give its analytic dependence on replication index $\nu$. Second, representation~\eqref{eq:FE-averaged-finite} holds only for integer numbers $\nu$. The replicated spins do not have a direct physical meaning and were introduced only to simulate scalings of the phase space to test thermodynamic homogeneity of the resulting free energy.  Hence, one has to continue analytically free energy~\eqref{eq:FE-averaged-finite} to a positive real replication index $\nu$. A specific symmetry of the matrix of the overlap susceptibilities must be assumed to reach this objective. Such a symmetry was found by G. Parisi \index{Parisi!Giorgio} and is independent of the replica trick and the limit of the number of replicas to zero.

\subsection{Hierarchical construction of mean-field free energies}

First what one notices is that the replicated free energy, Eq.~\eqref{eq:FE-averaged-finite}, contains too many order parameters, $\nu(\nu -1)/2$. It means that the solution is degenerate. Actually, the number of independent parameters in the $\nu$-times replicated phase space should not be bigger than the number of replicas, $\nu$. Parisi assumed the following structure of the overlap susceptibilities    
\begin{equation}\label{eq:chi-hierarchy}
\chi^{aa}= 0 \ , \quad \chi^{ab} = \chi^{ba} \ , \quad \sum_{c}\left(\chi^{ac} - \chi^{bc} \right) = 0 \ .
\end{equation}
It means that each row/column contains the same elements, only in a different order. This is a consequence of the third condition in Eq.~\eqref{eq:chi-hierarchy}. Hence the matrix of the overlap susceptibilities has just $\nu -1$ independent numbers. The values of the overlap susceptibilities can appear multiple times in each row. Let us assume that we have just $K$ different values of the overlap susceptibilities that we denote $\chi_{1}, \chi_{2},\ldots, \chi_{K}$. Let the  corresponding multiplicities be $m_{1}, m_{2}, \ldots,m_{K}$. A sum rule $\nu - 1 = \sum_{l=1}^{K}m_{l}$ then holds.  An example of such a matrix for $\nu = 8$, $K= 3$  and $m_{l}= 2^{l-1}$ is illustrated in Fig.~\ref{fig:hierarchy}.  \index{replica!symmetry breaking}

\begin{figure}\begin{center}
$$\begin{pmatrix}
0& \chi_1&
\chi_2&\chi_2&
\chi_3&\chi_3&
\chi_3 &\chi_3 \\
\chi_1
&0&\chi_2&\chi_2&
\chi_3&\chi_3&
\chi_3 &\chi_3  \\
\chi_2&\chi_2&
{0}&\chi_1&
\chi_3&\chi_3&
\chi_3&\chi_3  \\
\chi_2&\chi_2&
\chi_1& {0}&
\chi_3&\chi_3&
\chi_3&\chi_3   \\
\chi_3&\chi_3
&\chi_3&\chi_3&
{0}& \chi_1&
\chi_2&\chi_2  \\
\chi_3&\chi_3&
\chi_3&\chi_3&
\chi_1& {0}&
\chi_2&\chi_2  \\
\chi_3&\chi_3&
\chi_3&\chi_3&
\chi_2&\chi_2&
{0}& \chi_1  \\
\chi_3&\chi_3&
\chi_3&\chi_3&
\chi_2&\chi_2&
\chi_1& {0}
\end{pmatrix}$$
\caption{Matrix of overlap susceptibilities $\chi^{ab}$ for $\nu=8$ and with three levels (hierarchies) of symmetry breaking $K=3$ exemplifying the structure allowing for analytic continuation to arbitrary positive $\nu$. \label{fig:hierarchy}}
\end{center}\end{figure}

It is now straightforward to perform the sum over the spin configurations and find a closed expression for the free energy with matrices $\chi^{ab}$ fulfilling criteria~\eqref{eq:chi-hierarchy}. We label the overlap susceptibilities so that they form a decreasing succession  $\chi_{l}> \chi_{l+1}$.  The averaging over the thermal fluctuations of the replicated spins in the free energy from Eq.~\eqref{eq:FE-averaged-finite} can now be performed explicitly and we obtain \cite{Janis:2005aa}
\begin{subequations}\label{eq:f-discrete}
 \begin{multline}\label{eq:avfe-density}
    f_K(q,\{\chi\};\{m\}) = -\frac\beta 4 (1-q)^2 + \frac \beta 4
    \sum_{l=1}^K (m_l-m_{l-1})\chi_l(2q + \chi_l) + \frac \beta2
    \chi_1\\ -\ \frac 1{\beta m_{K}} \int_{-\infty}^{\infty} \frac
    {d\eta}{\sqrt{2\pi}} e^{-\eta^2/2}
    \ln\left[\int_{-\infty}^{\infty}\frac {d\lambda_K}{\sqrt{2\pi}}
      e^{-\lambda_K^2/2}\left\{\dots \int_{-\infty}^{\infty}\frac
        {d\lambda_1}{\sqrt{2\pi}}
        e^{-\lambda_1^2/2}\right.\right. \\ \left. \left.
        \left\{2\cosh\left[\beta\left(h +
              \eta\sqrt{q} + \sum_{l=1}^{K}\lambda_l \sqrt{\chi_l -
                \chi_{l+1}}\right)\right]\right\}^{m_1}\ldots\right\}
      ^{m_K/m_{K-1}}\right] \end{multline}
with $\chi_{K+1}=0$ and $m_{0} = 1$. It may appear convenient to rewrite the free-energy density to another equivalent form
 \begin{multline}\label{eq:mf-avfe}
f_K(q;\Delta\chi_1,\ldots,\Delta \chi_K, m_1,\ldots,m_K) =
-\frac\beta 4 \left(1-q -\sum_{l=1}^K\Delta\chi_l\right)^2 - \frac 1\beta
\ln 2 \\  + \frac \beta 4 \sum_{l=1}^K
m_l\Delta\chi_l\left[2\left(q + \sum_{i=l}^{K}\Delta\chi_{i}\right) -
\Delta\chi_l\right] - \frac 1\beta \int_{-\infty}^{\infty}
\mathcal{D}\eta\     \ln\   Z_K  
\end{multline}
\end{subequations}
where we ordered the parameters so that 
$\Delta\chi_l = \chi_l-\chi_{l+1}\ge\Delta\chi_{l+1}\ge 0$. We further used a short-hand notation for iterative partition functions
 $$ 
  Z_l =
\left[\int_{-\infty}^{\infty}\mathcal{D}\lambda_l\
Z_{l-1}^{m_l}\right]^{1/m_l} 
$$
with an abbreviation for a Gaussian differential
$\mathcal{D}\lambda \equiv {\rm d}\lambda\ e^{-\lambda^2/2}/\sqrt{2\pi}$. 
The initial partition function for the Ising spin glass is 
 $Z_0 =
\cosh\left[\beta\left(h + \eta\sqrt{q} + \sum_{l=1}^{K}\lambda_l
\sqrt{\Delta\chi_l} \right)\right]
$.
Free energy $f_{K}$ is an analytic function of multiplicities (geometric parameters) $m_{l}$  and hence they can now be arbitrary positive numbers. The equilibrium state in the replicated phase space is determined from the extremal point with respect to variations of the overlap susceptibilities, the order parameters in the glassy phase. With the  symmetry from Eq.~\eqref{eq:chi-hierarchy} the order parameters are the independent values of the overlap susceptibilities  $\chi_{l}$  in representation~\eqref{eq:avfe-density} or their differences $\Delta\chi_{l}$ from~\eqref{eq:mf-avfe} and their multiplicities $m_{l}$. The equilibrium values are determined from the extremal point of the respective free energy functional. The type of the extremum from which the stable equilibrium state is determined depends, however, on the values of multiplicities $m_{l}$. If $m_{l}>1$ then the equilibrium free energy is in minimum with respect to variations of this parameter. If $m_{l}<1$  then the equilibrium free energy is \textit{maximal}. The value $m_{l}=1$ is a degeneracy point at which the free energy is independent of $\Delta\chi_{l}$. It appears that the stable solution is generated by multiplicities being all between zero and one. The free energy of the spin glass is hence maximal \index{free energy!maximal} in the replicated phase space.  

Each independent value of the overlap susceptibility, $\chi_{l}$, determines a replica hierarchy. \index{replica!hierarchy} Here $K$ is the number of replica hierarchies. The number of replica hierarchies is also related to the level of the replica-symmetry breaking (RSB). \index{replica!symmetry breaking} The concept of replica-symmetry breaking comes from the replica trick \index{replica!trick} used to average free energy via the replicated partition function, Eq.~\eqref{eq:replica-trick}, where the order parameters in the replicated space are $q + \chi_{l}$. The number of the order parameters is $2K+1$. If all $\chi_{l}=0$ and $K=0$ we have a single order parameter $q$ and the solution is called replica-symmetric. It is the SK solution. Then non-zero values of $K$ are called $K$-level replica-symmetry breaking ($K$RSB). 

It is clear that the complexity of the solution increases rapidly with the increasing number of different values of of the overlap susceptibilities or their differences $\Delta\chi_{i}$, that is, with number $K$.  We give here an example of the lowest replica-symmetry breaking free energy ($K=1$)  as the next step beyond the SK solution
\begin{multline}\label{eq:FE-1RSB}
f_{1}(q;\chi_{1},m_{1}) = -\frac \beta4(1-q - \chi_{1})^2  + \frac \beta4m_{1}\chi_{1}(2 q + \chi_{1})  \\ -\frac 1{\beta m_{1}}
\int_{-\infty}^{\infty}\mathcal{D}\eta\ln\int_{-\infty}^{\infty}
\mathcal{D}\lambda_{1}  \left\{2\cosh\left[\beta\left(h +
\eta\sqrt{q} +\lambda_{1}\sqrt{\chi_{1}}\right)\right]\right\}^{m_{1}}\ .
\end{multline}
It has three parameters, $q,\chi_{1}, m_{1}$ to be determined from stationarity of the free-energy functional from Eq.~\eqref{eq:FE-1RSB}. It represents a free energy with the first level of ergodicity breaking or replica-symmetry breaking (1RSB). 

Generally, free energy $f_{K}$ stands for ergodicity breaking on $K$ levels, $K$ generations of replicas. When the replica symmetry is broken, it also means that ergodicity is broken and the thermodynamic limit of the original system depends on the behavior of the spins of the surrounding bath, being the replicated spins. The physical interpretation of breaking replica symmetry  is ergodicity breaking. The hierarchical replications of the system variables is then an iterative way to restore ergodicity or thermodynamic homogeneity in a larger phase space. The order parameters from the replicated space $m_{l},\Delta\chi_{l}$ then play the role of Legendre conjugate parameters controlling the energy exchange between the original system and its simulated thermal bath.      

Free energy $f_K(q;\Delta\chi_1,\ldots,\Delta \chi_K, m_1,\ldots,m_K) $ contains $2K + 1$ variational parameters, $q, \Delta\chi_{i}$, $m_{i}$ for $i=1,2,\ldots K$ that are determined from the stationarity of free energy with respect to small fluctuations of these parameters. The replica construction introduced a new parameter $K$ that is not \'a priori determined. It can assume any integer value in the true equilibrium. The number of replica hierarchies is in this construction determined from stability conditions that restrict admissible solutions, stationarity points. A solution with $K$ levels is locally stable if it does not decay into a solution with $K+1$ hierarchies. A new order parameter in the next replica generation \index{replica!generations}
$\Delta\chi$ may emerge so that $\Delta\chi_l > \Delta\chi >
\Delta\chi_{l+1}$ for arbitrary $l$. That is, the new order parameter may peel
off from $\Delta\chi_l$ and shifts the numeration of the order parameters
for $i>l$ in the existing $K$-level solution.  To guarantee that this does
not happen and that the averaged free energy depends on no more geometric
parameters than $m_1,\ldots,m_K$ we have to fulfill a set of $K+ 1$
generalized stability criteria \index{stability!criteria} that for our hierarchical solution
read for $l=0,1,\ldots,K$
\begin{equation}\label{eq:AT-hierarchical}
 \Lambda^{K}_{l} = 1 - \beta^2\left\langle\left\langle \left\langle 1 -
   t^2 + \sum_{i=0}^{l} m_i \left(\langle
    t\rangle_{i-1}^2 - \langle t\rangle_i^2\right)\right\rangle_{l}^2
  \right\rangle_K\right\rangle_\eta\ge 0 
\end{equation}
with $m_{0}= 0$ and formally $\left\langle t\right\rangle_{-1}=0$. 
We introduced the following
short-hand notation\\ $t \equiv \tanh\left[\beta\left(h + \eta\sqrt{q} + \sum_{l=1}^K
    \lambda_l\sqrt{\Delta\chi_l} \right)\right]$ and $\langle
t\rangle_l(\eta;\lambda_K,\ldots,\lambda_{l+1}) =
\langle\rho_l\ldots\langle\rho_1 t \rangle_{\lambda_1} \ldots
\rangle_{\lambda_l}$ with $\langle X(\lambda_l) \rangle_{\lambda_l} =
 \int_{-\infty}^{\infty}\mathcal{D}\lambda_l\
 X(\lambda_l)$ and 
$
\rho_l   = Z_{l-1}^{m_l}/\langle Z_{l-1}^{m_l}\rangle_{\lambda_l}
$. 
The lowest $K$ for which all stability conditions, Eq.~\eqref{eq:AT-hierarchical}, are fulfilled is an allowed equilibrium state. It need not, however, be the true equilibrium state, since the stability conditions test only the local stability and cannot decide which of several extremal points is the true ground state. The stability conditions, Eq.~\eqref{eq:AT-hierarchical}, are necessary for the system to be thermodynamically homogeneous. They are, however, not sufficient to guarantee global thermodynamic homogeneity. Note that the stability conditions from Eq.~\eqref{eq:AT-hierarchical} guarantee only local homogeneity, since they hold for the optimal geometric parameters $m_{l}$ determined by the stationarity equations. The global thermodynamic homogeneity, Eq.~\eqref{eq:av-homogeneity}, would demand $\Lambda^{K}_{K} \ge 0$ for arbitrary positive $m_{K}$. This is generally valid if $\Delta\chi_{K}=0$.

Free energy with $K$ hierarchies of replicated spin variables, Eqs.~\eqref{eq:f-discrete}, was derived by the standard procedure utilizing ergodicity in the  extended, replicated phase space. Real replicas in the thermodynamic  approach, that is without the replica trick, were introduced to include  control over thermodynamic homogeneity of the equilibrium state. The necessity to continue analytically the free energy to arbitrary positive replication index forced us to introduce a specific structure to matrix $\chi^{ab}$ as exemplified in Fig.~\ref{fig:hierarchy}.  Since $\chi^{ab}\ge 0$ we can consider it as a distance between replicas $a$ and $b$. The reduction rules, Eq.~\eqref{eq:chi-hierarchy} lead to an \textit{ultrametric distance} in the space of replicas \cite{Mezard:1984aa}. An ultrametric space \index{ultrametric!space} is characterized by the existence of only equilateral or isosceles triangles. That is, for three replica indices $a,b,c$ either $\chi^{ab} = \chi^{ac}= \chi^{bc}$ or at least one of these equalities holds. Ultrametricity is generated by the hierarchical structure of successive replications used to derive free energy  $f_K(q,\{\chi\};\{m\})$.   

The construction of the order parameters in the replicated phase space suggests that overlap susceptibilities $\chi^{ab}$ measure the interaction strength with which different copies (replicas) of spins thermodynamically influence each other. That is,  thermal
averaging of one spin copy ($a$) depends on the values of spins of another
copy ($b$) if $\chi^{ab}>0$. We cannot separate these replicas although
only one spin replica represents the physical system under consideration.
The non-replicated original phase variables together with temperature and
the chemical potential are hence insufficient to describe entirely the
equilibrium thermodynamic states. To get rid of the dependence of
thermodynamic states on the boundary or initial  conditions  we have to average
over all initial/boundary values and external variables that influence the
thermodynamics of the investigated system. In the long-range, completely
connected models the degeneracy of the solutions of the mean-field equations
is reflected in the dependence on the initial spin configurations. We
simulated this dependence by self-consistent interactions between the original and replicated 
spin variables, where both spin species are subject to the same thermal equilibration.

To understand the role of the geometric parameters (replication indices) $m_{l}$ we look at the $l$th level of replication. It appears that when ergodicity and the linear response with respect to inter-replica interaction is broken the solution in the replicated space has the replication indices ordered in a decreasing succession $1> m_{1} > \ldots m_{K} \ge 0$.  It  allows us to give a straightforward interpretation of the successive replications. Let us take partition sum $Z_{l-1}$ and perform the next replication of the spin variables from $Z_{l-1}$. The averaged interaction strength of the original and the replicated spins is $\Delta\chi_{l}$ and the averaged number of the original spins affected by the interaction with the replicated spins is $m_{l}N$. The spins from the subspace with $l-1$ replica hierarchies are affected by thermal fluctuations of the spins from the next,  $l$th hierarchy.  The Gaussian spins from the $l$th hierarchy are $\lambda_{l}$ and their thermal fluctuations are represented by a Gaussian integral. The free-energy density of the systems with $l$ replica hierarchies then is  
\begin{equation*}
f_{l}(\overline{h}_l)= \frac{1}{m_l } \ln \int\mathcal{D}\lambda_l\ Z_{l-1}^{m_l}
\left(\beta, \overline{h}_l + \lambda_l \sqrt{\Delta\chi_l}\right) \ ,
\end{equation*} 
where $\overline{h}_l$ is an internal magnetic field. 

\begin{figure}
\begin{center}
\includegraphics[width=8cm]{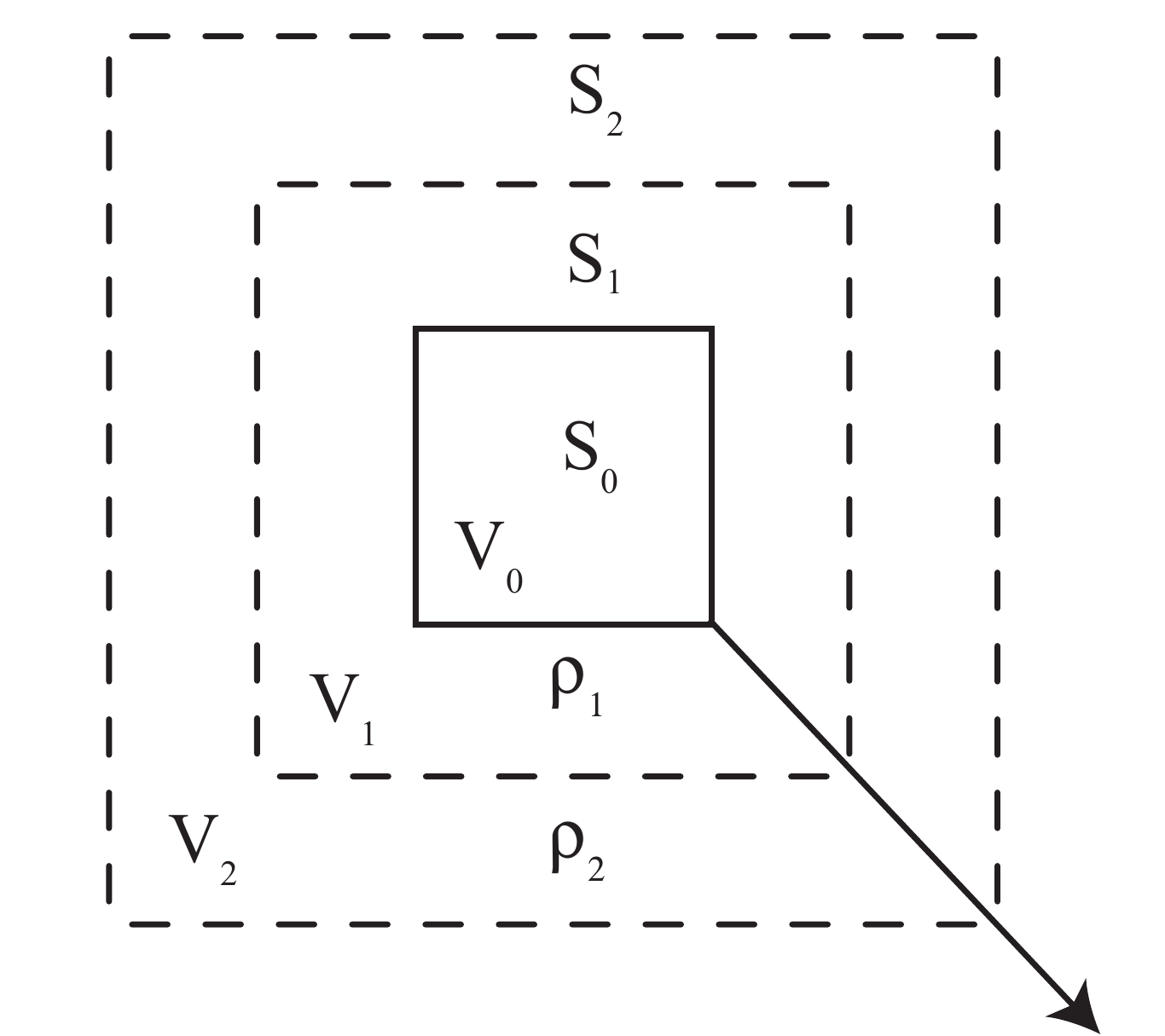}%
\caption{The thermodynamic limit of a system of spins $S_{0}$ confined in a volume $V_{0}\to\infty$ can either be performed in a free space or embedded in a bath of external spins $S_{1}$ in a larger volume $V_{1}\to\infty$. The bath spins have a density matrix $\rho_{1}$. If the two thermodynamic limits are different, we must embed the system into a next bath of spins with a density matrix $\rho_{2}$. We continue with the embeddings so long until the smaller and the larger system lead to the same thermodynamic limit.   \label{fig:emb} }%
\end{center}
\end{figure}

These hierarchical replications \index{replica!hierarchy} of the mean-field models can be understood as hierarchical embeddings of a finite volume $V$ into larger volumes with the surrounding spins, see Fig.~\ref{fig:emb}.  The density matrix of the spins from the $l$th  shell is  $\rho_l = Z_{l-1}^{m_l}/\langle Z_{l-1}^{m_l}\rangle_{\lambda_l}$. We can now perform the thermodynamic limit of the volume $V_{l-1} \to \infty$ with no surrounding spins or together with the next shell $V_{l} \to\infty$. If the the system in the volume $V_{l-1}$ is ergodic, the thermodynamic limit should not depend on the behavior of the surrounding spins. Here is the core of the problems and the instability of simple mean-field solutions of spin-glass models. They break ergodicity in the whole low-temperature phase and the replica-symmetry breaking is a mathematical representation of ergodicity breaking.  \index{ergodicity!breaking} The hierarchical construction is a way to incorporate the influence of the thermal bath on the original spin systems in the thermodynamic limit.  

\subsection{Continuous limit: Parisi solution}

Free energy $f_K(q;\Delta\chi_1,\ldots,\Delta \chi_K, m_1,\ldots,m_K)$ is an analytic function of its order parameters derived from $K$ hierarchical embeddings. Due to the embedded structure of the partition sums $Z_{l}$ it is practically impossible to find explicit solutions for $K>2$ in the whole temperature range of the glassy phase. Parisi found that in the case of the Ising spin glass solutions for $K=1,2$ are unstable and assumed that the same holds also for all finite $K$'s. \index{Parisi!solution} He then  performed the limit $K\to\infty$ and derived a continuous version of the infinitely-times replicated system by assuming 
$\Delta\chi_l = \Delta\chi/K \to dx$. Second and higher powers of $\Delta\chi_l$ with the fixed index~$l$ are  neglected \cite{Parisi:1980ab, Duplantier:1981aa,Parisi:1983aa}.  When performing the limit $K\to\infty$ in representations of Eqs.~\eqref{eq:f-discrete} the free-energy functional can then be represented as~\cite{Janis:2008aa, Janis:2013aa} \index{replica!symmetry breaking!continuous}
\begin{multline}\label{eq:FE-continuous} f(q,X; m(x)) = - \frac
  \beta4 (1 - q -X)^2 - \frac 1\beta \ln 2\\ + \frac {\beta }2
  \int_0^Xdx\ m(x)\left[q + X - x \right] - \frac
  1\beta \left\langle g(X, h + \eta \sqrt{q})\right\rangle_\eta
\end{multline} %
where $\langle X(\eta)\rangle_\eta = \int_{-\infty}^\infty
\mathcal{D}\eta X(\eta)$. This free energy is only implicit since its interacting part $g(X,h)$ can be expressed only via an integral representation containing the solution itself 
\begin{subequations}
\begin{align}\label{eq:g0}
  g(X,h) &= \mathbb E_0(h;X,0)\circ g_0(h) \nonumber\\ & \equiv
  \mathbb T_x \exp\left\{\frac 12 \int_0^X dx
    \left[\partial_{\bar{h}}^2 
   + m(x) g'(x;h + \bar{h})\partial_{\bar{h}}
    \right] \right\} g(h + \bar{h})\bigg|_{\bar{h}=0}\ ,
\end{align}
 with $g(h) = \ln \left[\cosh\beta h\right]$.  The "time-ordering" operator \index{operator!time ordering} $\mathbb T_x$ orders products of $x$-dependent non-commuting operators from left to right in an  $x$-decreasing succession. The exponent of the ordered exponential contains function $g'(x;h)= \partial g(x;h)/\partial h$ for $x\in [0,X]$ and is not known when $g(x;h)$ is not know on the whole definition interval. This derivative can also be expressed via an ordered exponential 
\begin{align}\label{eq:g1}
  g'(X,h) &= \mathbb E(h;X,0)\circ g_0^\prime(h) \nonumber \\
  &\equiv \mathbb T_x \exp\left\{  \int_0^X dx
    \left[\frac 12 \partial_{\bar{h}}^2 + m(x) g'(x;h +
      \bar{h})\partial_{\bar{h}} \right] \right\} g_0'(h +
  \bar{h})\bigg|_{\bar{h}=0}\ .
\end{align}
\end{subequations}
It is an implicit but closed functional equation for the derivative $g^{\prime}(x;h)$ on interval $[0,X]$ for a given function $m(x)$. We have to know the full dependence of this function on parameter $x$ to evaluate the free energy with continuous replica-symmetry breaking. \index{replica!symmetry breaking!continuous} It is important to note that free energy $f(q,X; m(x))$ defines a thermodynamic theory independently of the replica method within which it was derived. It means that we can look for equilibrium states of spin-glass models without the necessity to go through instabilities of the discrete hierarchical replica-symmetry breaking solutions.  

Analogously we can perform the limit $K\to\infty$ with $m_{l} - m_{l-1} = -\Delta m /K \to - dm$ in representation Eq.~\eqref{eq:avfe-density}.  The minus sign is used, since we expect ordering $m_{l}< m_{l-1}$. Further on,  we use $\Delta\chi_{l} \to -x(m) dm$. The limiting continuous free energy can then be represented as \cite{Janis:2015aa}
\begin{multline}\label{eq:FE-mcontinuous} f\left(q,\chi_{1}, m_{1},m_{0}; x(m)\right) = - \frac\beta4 (1 - q -\chi_{1})^2 + \frac\beta 4 \left[m_{1} \left( q + \chi_{1}\right)^{2} - m_{0} q^{2}\right] 
\\  -  \frac {\beta}4
  \int_{m_{0}}^{m_{1}}dm\left[q + \chi_{1} - X(m)\right]^{2} -  \frac 1\beta \left\langle g_{1}(m_{0}, h + \eta \sqrt{q})\right\rangle_\eta \ ,
\end{multline} %
where we denoted $X(m) = \int_{m}^{m_{1}}dm'x(m')$ and 
\begin{subequations}
\begin{multline}\label{eq:gm1}
  g_{1}(m_{0}, h) = \mathbb E_0(m_{0},m_{1},h)\circ g_1(h) \\ \equiv
  \mathbb T_m \exp\left\{-\frac 12 \int_{m_{0}}^{m_{1}} dm \ x(m)
    \left[ \partial_{\bar{h}}^2 
   +  m g'_{1}(m;h + \bar{h})\partial_{\bar{h}}
    \right] \right\} g_1(h + \bar{h})\bigg|_{\bar{h}=0}\ ,
\end{multline}
and
\begin{equation}\label{eq:g_1-Zero}
g_{1}(h) \equiv  g_{1}(m_{1},h) = \frac 1{m_{1}} \ln \int_{-\infty}^{\infty}\frac {d \phi}{\sqrt{2\pi}}  e^{-\phi^{2}/2} \left[ 2\cosh\left(\beta(h + \phi\sqrt{\chi_{1}})\right) \right]^{m_{1}}\ .
\end{equation}\end{subequations}
This free energy better suits the case when the solution with continuous RSB peels off from  a solution with one-level RSB or the two solutions  coexist. The space of order parameters is restricted to an interval $1\ge m_{1}\ge m_{0}$ and $1 \ge \chi_{1} \ge X(m) \ge 0$.

If free energy $f_{K}$ is not locally stable and at least one of the stability conditions, Eq.~\eqref{eq:AT-hierarchical}, is broken for all $K$'s, it is still a question whether the continuous limit $K\to\infty$   is locally stable. It can be shown that continuous free energy $f(q,X; m(x))$ is marginally stable \index{stability!marginal} and fulfills the continuous version of stability conditions with equality~\cite{Janis:2008aa}
\begin{equation}\label{eq:marginal_{stability}}
1 =  \left\langle \mathbb E(h_\eta;X,x)\circ\left[ g_{\mu}^{\prime\prime}(x,h_{\eta})^{2}\right]\right\rangle_{\eta}\ .
\end{equation}
This equation is a consequence of the stationarity equation for the order-parameter function $m(x)$.  We recall that the prime stands for the derivative with respect to the magnetic field and the second derivative of $ g_{\mu}(x,h)$ obeys an integral equation 
\begin{equation}\label{eq:gx1}
  \frac {\partial^{2} g_{\mu}(x,h)}{\partial h^{2}} = \mathbb
  E(h;x,0)\circ g_{\mu}^{\prime\prime}(h) +  \int_0^x
  d y\ \mu(y)\mathbb E(h;x,y)\circ \left[g_{\mu}^{\prime\prime} (y,
    h)^{2}\right]\ .
\end{equation}
 The continuous free energy hence does not break ergodicity and is always marginally ergodic in the whole spin-glass phase.  The ferromagnetic model is marginally ergodic only at the critical point, since the order parameter makes the ordered phase ergodic.     

Both continuous free energies, Eq.~\eqref{eq:FE-continuous} and~\eqref{eq:FE-mcontinuous},  were derived as the limit of the number of replica hierarchies $K\to\infty$ where the distance between the neighboring hierarchies is infinitesimal, that is $\Delta\chi_{l} \propto K^{-1}$, and $\Delta\chi_{l}/\Delta m_{l}< \infty$ for each $l\le K$. Representation~\eqref{eq:FE-mcontinuous} is, however, more general, since it does not assume $\chi_{K}\to 0$ in the limit $K\to\infty$ as free energy, Eq.~\eqref{eq:FE-continuous}, does. On the other hand, condition $\chi_{K}\to 0$ guarantees thermodynamic homogeneity of the resulting free energy.  

The continuous free energies were derived for the Ising spin glass but they can be straightforwardly generalized to other spin-glass  models. The symmetry of the order parameters has to be adapted and the input single-site free energy $g$  or $g_{1}$ is to be appropriately modified~\cite{Janis:2011ab,Janis:2013aa}.

\section{Asymptotic solutions of mean-field models: K-level replica symmetry breaking}

\index{solution!asymptotic}

\subsection{Ising glass}

The Sherrington-Kirkpatrick model is the paradigm for the mean-field theory of spin glasses. It is this model for which Parisi derived a free energy with  continuous replica-symmetry breaking~\cite{Parisi:1980ac,Parisi:1980ab,Parisi:1980ae}. It was also later proved that the hierarchical scheme of replica-symmetry breaking covers the exact equilibrium state~\cite{Guerra:2003aa,Talagrand:2006aa}.  The rigorous proof does not, however, tell us whether the equilibrium state is described only by a finite number of replica hierarchies  or a continuous limit is needed. Only a few years ago we resolved the hierarchical free energy $f_K(q;\Delta\chi_1,\ldots,\Delta \chi_K, m_1,\ldots,m_K)$ for arbitrary $K$ via the asymptotic expansion below the transition temperature in a small parameter $\theta = 1 - T/T_{c}$~\cite{Janis:2006ac}. Only this asymptotic solution was able to resolve the question of the structure of the equilibrium state, at least close to the transition temperature. \index{mean field model!Ising glass}

\begin{figure}\begin{center}
\includegraphics[width=11cm]
{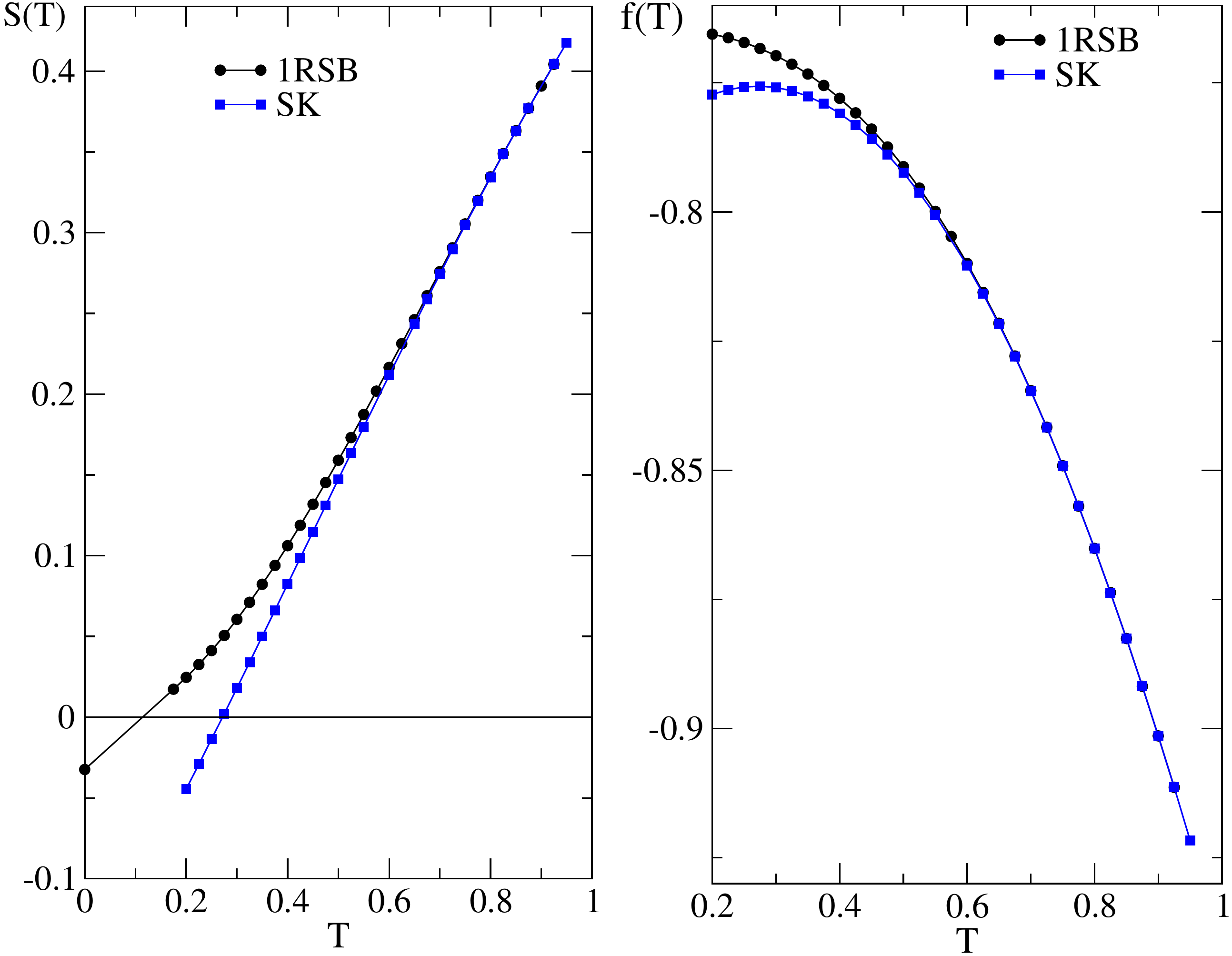}
\caption{Entropy (left panel) and free energy (right panel) in the glassy phase of the Sherrington-Kirkpatrick model at zero magnetic field for the replica-symmetric and 1RSB solutions. Improvements at very low temperatures are evident, in particular, the unphysical negative value of entropy is significantly reduced. \label{fig:1RSB-Ising}}\end{center}
\end{figure}

\subsubsection{Zero magnetic field}

The order parameters in the spin-glass phase of the general $K$-level
free energy from Eqs.~\eqref{eq:f-discrete} cannot be solved explicitly
in the whole temperature range unless we resort to solutions with only a few hierarchical levels
($K=1,2$). The replica symmetric and 1RSB solutions in the whole low-temperature phase are plotted in Fig.~\ref{fig:1RSB-Ising} where one can see the improvements of the replica-symmetry breaking scheme in free energy and entropy. One has to go to higher replica hierarchies. The only chance to analyze the behavior of the entire hierarchical construction with an arbitrary number of hierarchies is to expand the solution near the critical point where the order parameters are small.

The strategy to solve the stationarity equations for the hierarchical
free energy asymptotically near the critical temperature is to expand
the partition function into powers of the small order parameters and
to restrict the solution only to a functional subspace generated by a
fixed polynomial expansion. We first use such an expansion to derive
the leading asymptotic limit of equations for the physical parameters
$q, \Delta\chi_l$. At this stage we do not need to assume smallness of
the geometric parameters. Smallness of $m_l, l=1,\ldots, K$ at zero
magnetic field is utilized later on when deriving the asymptotic
form of mean-field equations for them.

After a rather lengthy and tedious calculations one finds the leading order of the order parameters \cite{Janis:2006ac}
\begin{subequations}\label{eq:order-parameters}
\begin{align}\label{eq:chi_i}
  \Delta\chi_l^K &\doteq \frac 2{2 K + 1}\ \theta\ ,\\ \label{eq:m_i}
  m_l^K &\doteq \frac {4 (K - l + 1)}{2 K + 1}\ \theta \ ,\\ \label{eq:q}
  q^K &\doteq \frac 1{2 K + 1}\ \theta \ .
\end{align}
The result proves that the limit $K\to\infty$ leads indeed to the Parisi continuous replica-symmetry breaking, since all the order parameters are of order $K^{-1}$.  Each solution with a finite number of replica generations is unstable
\begin{equation}\label{eq:Lambda-solution}
 \Lambda_l^{K} = -\frac  43 \ \frac {\theta^2}{(2 K + 1)^2} < 0\ ,
\end{equation}
\end{subequations}
from which it follows that the Parisi solution is the true equilibrium state in the glassy phase of the Sherrington-Kirkpatrick model near the transition point. 

Other physical quantities in the asymptotic limit are the Edwards-Anderson parameter  \index{Edwards-Anderson!parameter} defined as $Q^{K}= q + \sum_{l=1}^{K}\Delta \chi_{l}^{K}$ with its asymptotics
\begin{subequations}
\begin{align}\label{eq:Qprime}
  Q^K &\doteq \theta + \frac {12 K(K + 1) + 1}{3(2 K +1)^2}\ \theta^2\  ,
\end{align}
the local spin susceptibility
\begin{equation}\label{eq:thermal-susceptibility}
\chi_T = \beta \left(1 - Q^{K} + \sum_{l=1}^K m_l\Delta\chi_l\right) \doteq 1 -
\frac{\theta^2} {3(2K + 1)^2}\ . 
\end{equation}
and the free-energy difference to the paramagnetic state 
\begin{equation}\label{eq:FE-better-asymptotic}
\Delta f\doteq\left(\frac{1}{6}\theta^3+\frac{7}{24}
\theta^4+\frac{29}{120}\theta^5\right)-\frac{1}{360} \theta^5
   \left(\frac{1}{K}\right)^4 \ .
\end{equation}
\end{subequations}
Differences between different levels of RSB manifest themselves  in free energy first in fifth order to which one had to expand the free energy in the order parameters.  

\subsubsection{Non-zero magnetic field} 

The glassy phase in the Sherrington-Kirkpatrick model exists also in an arbitrary strong applied magnetic field. The transition boundary separating the paramagnetic from the glassy phase is the  de Almeida-Thouless  line. It is defined by vanishing of the stability function $\Lambda=0$ of Eq.~\eqref{eq:AT-condition}.  The hierarchical solution behaves in an external magnetic field differently from the rotationally invariant case. The physical order parameter $q$ and the geometric order parameters $m_l$ remain finite at the transition to the paramagnetic phase and only the differences of the overlap susceptibilities $\Delta\chi_l$ are the genuine small parameters controlling the expansion around the critical point. A small expansion parameter below the AT line can be chosen as
\begin{equation}\label{eq:alpha-def}
  \alpha = \beta^2\left\langle (1 - t_0^2)^2\right\rangle_\eta  - 1 > 0\ ,
\end{equation}
where we denoted $t_{0} = \tanh\beta\left(h + \eta \sqrt{q_{0}}\right)$,  and $q_{0}=\left\langle t_{0}^{2}\right\rangle_{\eta}$.

One must first expand the SK parameter $q$ to the two lowest nontrivial
orders in $\alpha$. The solution is then used to determine the lowest
asymptotic order of $ \chi_l$ and $m_l$.  We obtain $m_l = m_{1} +
O(\alpha)$  and $\chi_{1}= \sum_{l=1}^{K}\Delta\chi_{l}^{K}$ with\cite{Janis:2008ac}
\begin{equation}\label{eq:m0-solution}
  m_{1} =  \frac {2\langle t_0^2(1 - t_0^2)^2\rangle_{\eta}}{\langle (1
    -  t_0^2)^3\rangle_{\eta}}
\end{equation}
and
\begin{eqnarray}\label{eq:X-solution}
  \chi_1 = \frac{\alpha}{2\beta^2m_{1}} \ \frac 1{1 -
      3\beta^2\left\langle t_0^2(1 -  t_0^2)^2\right\rangle_\eta} +
O(\alpha^2)\  \ . \end{eqnarray}
These two parameters do not depend on the number of hierarchical
levels used.  Parameter $m_{1}$ is of order unity even at the boundary of the
spin-glass phase (AT line) where the small parameter $\alpha$
vanishes.

To disclose the leading asymptotic behavior of each separate parameter
$\Delta\chi_l$ and $ m_l$ for $l=1,\ldots, K$ we must go beyond
 the leading orders in parameters $m$ and $ \chi_1$. It is
first the fourth order in $\alpha$  from which we find that $\Delta\chi_l \doteq \chi_1/K$ and
\begin{equation}\label{eq:ml}
m^K_l \doteq m_{1} + \frac {K + 1 - 2l}K\ \Delta m
\end{equation}
where we added a superscript to specify the number of hierarchical levels
used to determine the order parameters $\chi_l,m_l$. Further on, we
introduced a parameter independent of the number of hierarchies  $\Delta m
=m_1^2 - m_2^2$.  This parameter has an explicit asymptotic representation
\begin{equation}\label{eq:M-solution} \Delta m \doteq \frac
  {\beta ^2
    \chi_1\left\langle \left(1 - t_0^2\right)^2 \left(2 \left(1-3
          t_0^2\right)^2+3 \left(t_0^2-1\right) m \left(8
          t_0^2+\left(t_0^2-1\right) m \right)\right)
    \right\rangle_\eta}{\left\langle (1 - t_0^2)^3 \right\rangle_\eta}
\end{equation}

Both parameters $ \chi_1$ and $\Delta m$ are linearly proportional to
$\alpha$. The former, however, exists already in $1$RSB, while the
latter first emerges in $2$RSB. Since they do not depend on the number
of hierarchies used and the latter determines a uniform distribution
of parameters $m_l$ for $l = 3,\ldots,K$, we demonstrated that all
characteristic features of the asymptotic solution near the AT
instability line are contained already in $2$RSB. What was, however,
highly nontrivial was to unveil equidistant distributions of both
parameters $ \chi_l$ and $ m_l$. 

The stability conditions 
\begin{equation}\label{eq:Lambda_K}
  \Lambda_K = - \ \frac{2\beta^2}{3 K^2} \ \frac{ \chi_1 \Delta m}{m +
    2 }\end{equation}
indicate that also in an applied magnetic field the equilibrium state is described by the continuous replica-symmetry breaking $K\to\infty$.

\subsection{Potts glass}

The Potts model with $p$ states reduces to the Ising model for $p=2$, but differs from it for $p>2$ in that it breaks the spin-reflection symmetry. This property was used to argue that the Parisi scheme fails to describe the equilibrium state~\cite{Goldbart:1985aa}. It had been long believed that it is the one-level replica-symmetry breaking that determines the equilibrium state below the transition temperature~\cite{Cwilich:1989aa}. The Potts glass displays a discontinuous transition \index{transition!discontinuous} to the replica-symmetry broken state for $p>4$~\cite{Gross:1985aa}. Discontinuous transitions do not allow us to use an asymptotic expansion in a small parameter below the transition temperature. It is, nevertheless, possible to test the ordered phase of the Potts glass for $2< p< 4$. We did it in Refs.~\cite{Janis:2011ab,Janis:2011aa} and found an unexpected behavior.   \index{mean field model!Potts glass} 

Studying the discrete replica-symmetry breaking we found two 1RSB solutions with the same geometric parameter  
\begin{align}
m &\doteq \frac{p-2}{2}+\frac{36-12 p +p^2}{8(4-p)}\theta\ .
\end{align}
One non-trivial 1RSB solution then leads to order parameters
\begin{subequations}\label{eq:1RSB-Sol1}
\begin{align}
q^{(1)}&\doteq 0\ ,\\
\Delta\chi^{(1)}&\doteq \frac{2}{4-p}\theta + \frac{228-96 p +p^2}{6 (4-p)^3}\theta^2
\end{align}
\end{subequations}
while the second one has both parameters nonzero
\begin{subequations}\label{eq:1RSB-Sol2}
\begin{align}\label{eq:1RSB-Solq2}
q^{(2)}&\doteq \frac{-12+24 p -7 p^2}{3(4-p)^2 (p-2)}\theta^2 \ ,\\
\Delta\chi^{(2)}&\doteq \frac{2}{4-p}\theta - \frac{360-204 p -6 p^2+13 p^3}{6 (4-p)^3 (p-2)}\theta^2 \ . \label{eq:1RSB-Solchi2}
\end{align}\end{subequations}
Both the solutions have the same asymptotic free energy to the fifth asymptotic order
  \begin{multline}\label{f:1RSB:tau} \frac{\beta}{p-1} f_{1RSB}\doteq
    \frac{\theta ^3}{3 (4-p)} + \frac{(p (11 p-102)+204) \theta ^4}{12 (4-p)^3} \\
    -\frac{(p (p ((18744-1103 p) p-120648)+325728)-317232) \theta ^5}{720
      (4-p)^5} \ . 
 \end{multline}
We can see that the asymptotic expansion with small parameters $q$ and $\Delta\chi$ breaks down already at $p=4$ above which we expect a discontinuous transition from the paramagnetic to a 1RSB state at $T_{0} > T_{c}=1$. Note that a transition to the replica-symmetric solution $q> 0,\Delta\chi =0$ is continuous up to $p=6$. 

The 1RSB solution has a higher free energy than the replica-symmetric one. The difference is of order $\theta^{3}$,  
\begin{equation}
f_{1RSB} - f_{RS} \doteq \frac {(p-2)^{2}(p-1) \theta^{3}}{3(4-p)(6-p)^{2}}\ .
\end{equation}
The two stationary states of the 1RSB free energy behave differently as a function of parameter $p$. The former solution is physical for all values of $p$ unlike the latter that becomes unphysical for $p>p^{*}\approx 2.82$ where $q^{(2)}$ from Eq.~\eqref{eq:1RSB-Solq2} turns negative. It is also the region of the parameter $p$ where the first solution is locally stable as can be seen from the stability function      
\begin{equation} \label{eq:1RSB-instability1}
\Lambda^{1}_{0} 
\doteq\frac{\theta^{2}(p - 1)}{6(4 - p)^{2}} \left(
    7p^{2} - 24 p + 12\right)  > 0\ .
\end{equation} 
That is why the solution with $q=0$ was assumed to be the true equilibrium and a solution with a continuous replica-symmetry breaking had not been expected to exist. We, however, found that there is a Parisi-like solution even in the region of stability of the solution from Eq.~\eqref{eq:1RSB-Sol1}. The second 1RSB solution is unstable and decays to solutions with higher numbers of replica hierarchies as 
\begin{subequations}
\begin{align}\label{eq:KRSB-q_As}
q^{K}&\doteq -\frac{1}{3K^{2}}\frac{12-24p+7p^{2}}{(4-p)^{2}(p-2)}\theta^{2}\ ,\\
\label{eq:KRSB-Dchi_As}
\Delta\chi_{l}^{K}&\doteq \frac{1}{K}\frac{2}{(4-p)}\theta \ ,
\\  \label{sol:chi}
m_{l}^{K}& \doteq\frac{p-2}{2}+
\frac{2}{4-p}\left[3+\frac{3}{2}p-p^{2}
+\left(3-6p+\frac{7}{4}p^{2}\right)\frac{2l-1}{2K}\right]\theta\ .
\end{align}
\end{subequations}
We can see that the $K$RSB solution behaves unphysically in the same way as the second 1RSB solution does. The averaged square of the local magnetization is negative for $p > p^{*}$ where the first 1RSB solution is locally stable. Negativity of $q$ means that local magnetizations are  imaginary and the solution is unphysical. This deficiency, however, decreases with the increasing number of spin hierarchies and disappears in the limit $K \to \infty$. It means that the resulting solution with a continuous replica-symmetry breaking shows no unphysical behavior. It is analogous to negativity of entropy in the low-temperature solutions of $K$RSB approximations of the Sherrington-Kirkpatrick model.  

\begin{figure}\begin{center}
\includegraphics[width=8cm]{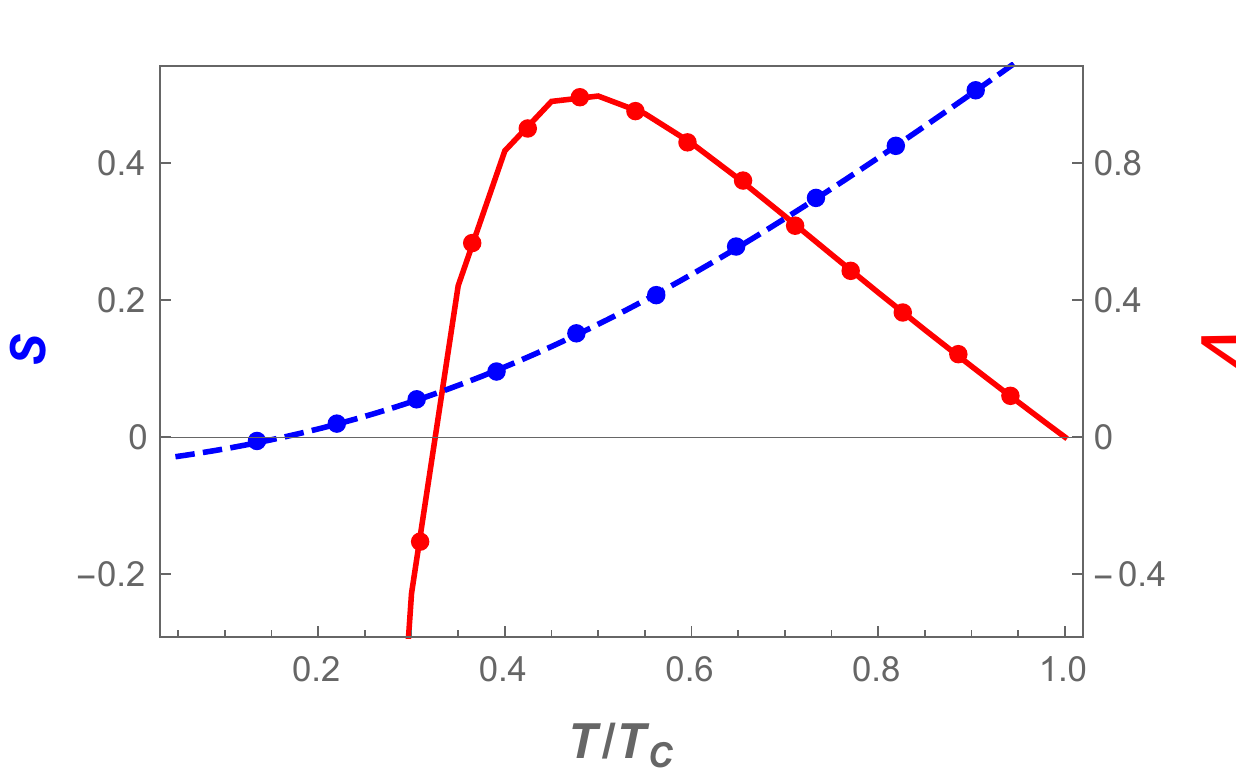}\includegraphics[width=8cm]
{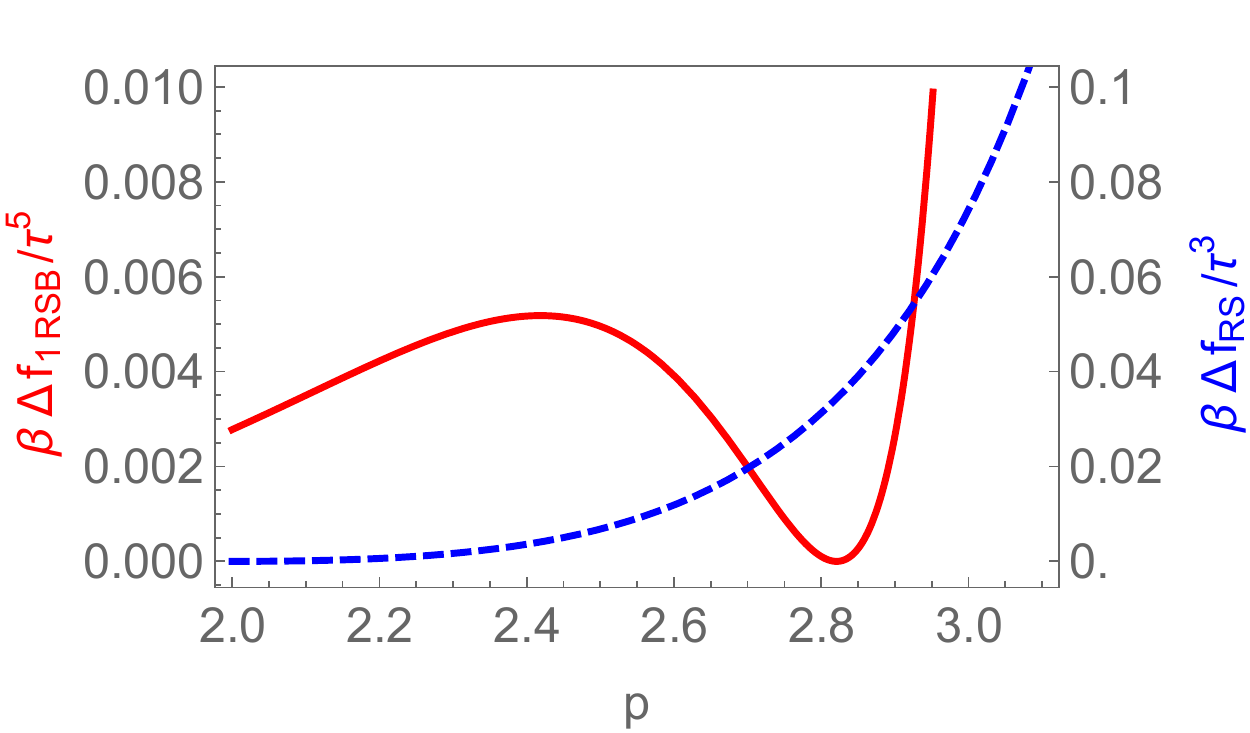}
\caption{Left panel: Entropy  $S$ (left scale, dashed line)  and local stability $\Lambda$ (right scale, solid line) of the 1RSB solution from Eq.~\eqref{eq:1RSB-Sol1} of the $3$-state Potts glass. The solution becomes locally unstable at $\approx 0.33 T_{c}$ and entropy negative at $T\approx 0.16 T_{c}$. Right panel: The free energy differences between the solution with continuous replica-symmetry breaking and 1RSB (left scale, solid line) and RS (right scale, dashed line) solutions. \label{fig:1RSB-Potts}}\end{center}
\end{figure}

Potts glass hence shows a degeneracy \index{solution!degenerate} for $p^{*}< p< 4$ with a marginally stable solution continuously breaking  the replica symmetry and a locally stable one-level replica-symmetry breaking. To decide which one is the true equilibrium state one has to compare free energies. The difference of the continuous free energy $f_{c}$ and that of the $K$RSB solution is
\begin{subequations}
\begin{equation}\label{eq:fdiff} \beta (f_{c}-f_{KRSB})\doteq
  \frac{(p-1) (p (7 p-24)+12)^2 \theta ^5}{720 K^{4}(4-p)^5} 
\end{equation} 
and that of the replica-symmetric one reads
\begin{equation}
  \beta(f_{c } - f_{RS}) \doteq \frac{(p - 1) (p - 2)^{2}\theta^{3}}{3(4 - p)(6
    - p)^{2}}\ .
\end{equation} 
\end{subequations}
We see that the solution with the continuous RSB has the highest free energy as the true equilibrium state should have for geometric factors $m<1$. In this situation entropy reaches minimum and free energy maximum in the phase space of the order parameters. The locally stable 1RSB solution becomes unstable at lower temperatures and entropy turns negative at very low temperatures as demonstrated on the $3$-state model in Fig.~\ref{fig:1RSB-Potts}. 
This leads us to the conclusion that the Parisi solution with a continuous replica-symmetry breaking represents the equilibrium state for the Potts glass with $p<4$.    

\subsection{$p$-spin glass}

The spin model generalized to random interactions connecting $p$ spins, the $p$-spin glass, was used to simulate the dynamical transition in real glasses~\cite{Kirkpatrick:1987aa,Kirkpatrick:1987ab}. This model, analogously to the Potts glass, generalizes the Ising spin glass to $p>2$ and allows one to study the behavior of the equilibrium state as a function of parameter $p$. In particular, the limit $p\to\infty$ is accessible~\cite{Gross:1984aa} and is exactly solvable. It coincides with the random energy model of Derrida~\cite{Derrida:1980aa,Derrida:1981aa}. For this reason the $p$-spin glass was also intended to be used to study and understand the genesis of the Parisi free energy when studying the asymptotic limit $p\to \infty$. \index{mean field model!p-spin glass}

To cover both the boundary solutions $p=2$ and $p=\infty$ we have to mix up the one-level RSB scheme and the Parisi continuous RSB. Such free energy density of the mean-field $p$-spin glass reads \cite{Janis:2015aa}
\begin{multline}\label{eq:FE-general}
f^{(p)}_{T}(q, \chi_{1},\mu_{1},\mu_{0}; x(\mu)) =  - \frac{\beta}{4}\left[ 1 - p\left(q + \chi_{1}\right) + (p-1) \left(q + \chi_{1}\right)^{p/(p-1)} \right]     \\
+ \frac{p-1}{4} \left[\mu_{1}\left(q + \chi_{1}\right)^{p/(p-1)} - \mu_{0}q^{p/(p-1)}\right] - \frac {p - 1} 4 \int_{\mu_{0}}^{\mu_{1}} d \mu \left[q +  \chi_{1} - X(\mu) \right]^{p/(p-1)}\\ -  g_{1}(\mu_{0},h)\ ,
\end{multline}
with 
\begin{subequations}
\begin{multline}\label{eq:g0x}
  g_{1}(\mu_{0},h) =  \mathbb E_1(\mu_{1},\mu_{0},h)\circ\left[ g_{1}(h)\right]  \equiv
  \mathbb T_{\mu} \exp\left\{- \frac p 4 \int_{\mu_{0}}^{\mu_{1}} d \mu x(\mu)
    \right. \\   \left. \left[\partial_{\bar{h}}^2  
 + \ \mu g_{1}^{\prime}(\mu,h + \bar{h})\partial_{\bar{h}}
    \right] \right . \bigg\} g_{1}(h + \bar{h})\bigg|_{\bar{h}=0} \ ,
\end{multline}
where  $X(\mu) =  \int_{\mu}^{\mu_{1}}d\mu' x(\mu')$. The generating free energy is 
\begin{equation}\label{eq:g_{mu}-Zero}
g_{1}(h) = \frac 1\mu_{1} \ln \int_{-\infty}^{\infty}\frac {d \phi}{\sqrt{2\pi}}  e^{-\phi^{2}/2} \left[ 2\cosh\left(\beta(h + \phi\sqrt{p\chi_{1}/2})\right) \right]^{\mu_{1}/\beta}\ .
\end{equation}
\end{subequations}
We rescaled variable  $m \to \mu = \beta m$.  If $\mu = 0$ or $\mu_{1}= \mu_{0}$, free energy $f^{(p)}$ reduces to the 1RSB approximation. On the other hand, if $\mu_{0}=0$, or $\mu_{0}= \beta$ free energy  $f^{(p)}$ coincides with that of the Parisi solution with a continuous replica-symmetry breaking. 

\begin{figure}\begin{center}
\includegraphics[width=11cm]{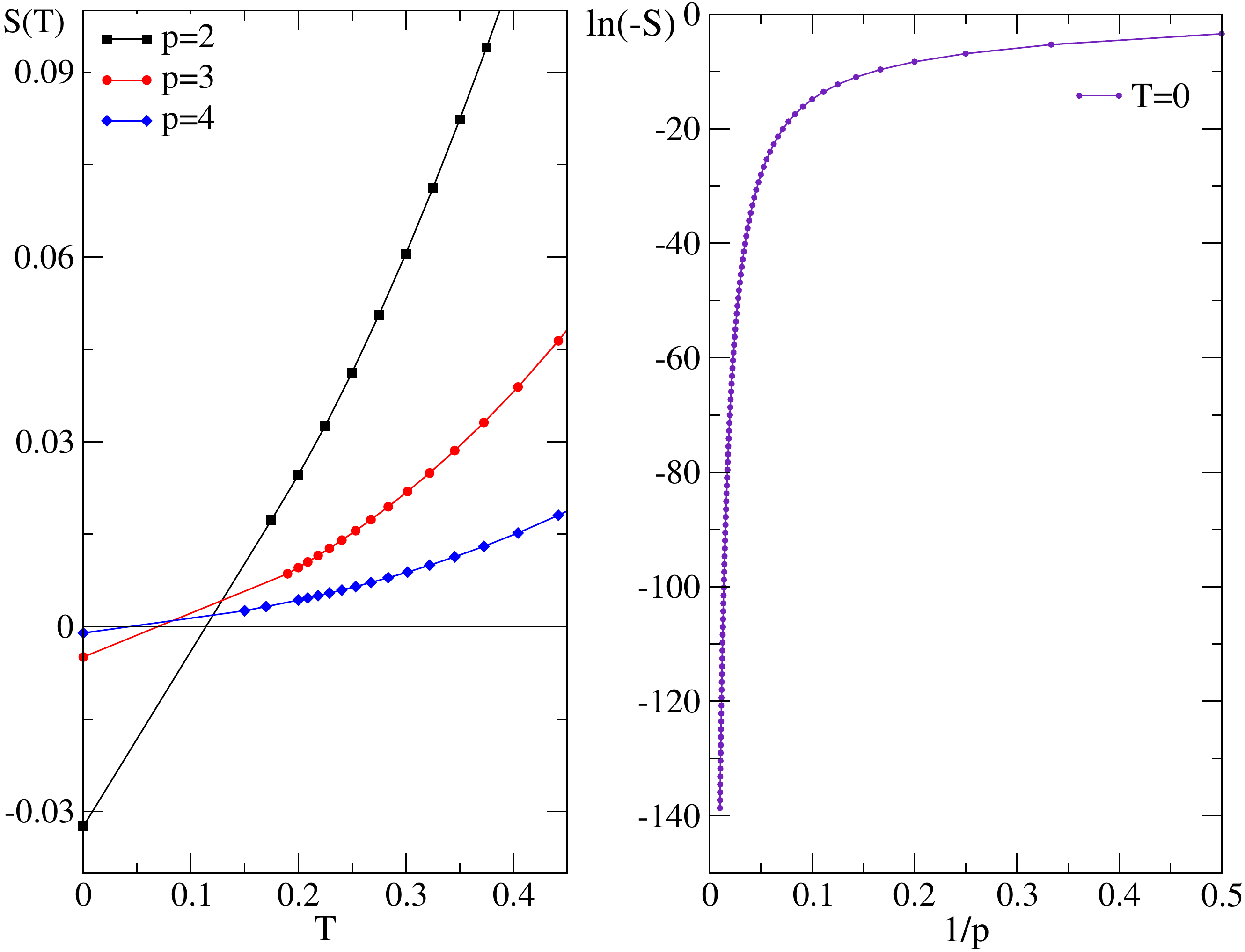}
\caption{Entropy of the 1RSB solution of the $p$-spin glass. Temperature dependence for  different values of $p$ (left panel). Logarithm of the negative part of entropy at zero temperature as a function of $1/p$  (right panel), from Ref.~\cite{Janis:2015aa}. \label{fig:1RSB-entropy}}\end{center}
\end{figure}

The $p$-spin glass can be used to investigate analytically not only the $p\to\infty$ limit but also the $T\to 0$ limit. In this limit simple solutions of mean-field models lead to negative entropy. It is easy to calculate the zero-temperature entropy in the 1RSB solution. We obtain
\begin{align}\label{eq:S-Tlow}
S_{0}(h) &\propto  - \frac{p(p-1)}{8} \left[\frac{\exp\{- \mu_{1}^{2}p\chi_{1}/4\}}{\sqrt{\pi p\chi_{1}}}\  \frac{\exp\{- h^{2}/p\chi_{1}\}}{2CH_{\mu}(h)}\right]^{2}  \ , 
\end{align}
where we used the following notation
\begin{subequations}
\begin{align}\label{eq:E-CH}
2CH_{\mu}(h) & = e^{\mu_{1}h} E_{\mu}^{(p)}(- h) + e^{-\mu_{1}h} E_{\mu}^{(p)}(h)  \ ,
\\ \label{eq:E-mu}
E_{\mu}^{(p)}(h) &= \int_{h/\sqrt{p\chi_{1}/2}}^{\infty} \frac{d \phi}{\sqrt{2 \pi}} e^{-\left(\phi - \mu_{1}\sqrt{p\chi_{1}/2}\right)^{2}/2} \ .
\end{align}
\end{subequations}
The negativity of the low-temperature entropy indicates that 1RSB cannot produce a stable ground state for arbitrary $p<\infty$. The negativity of entropy \index{entropy!negative} decreases with increasing $p$, see Fig.~\ref{fig:1RSB-entropy}, but only if a condition $\mu_{1}^{2} p \chi_{1} = \infty$ is fulfilled the 1RSB solution ($\mu_{1}>0$) leads to zero entropy at zero temperature. Nonnegative entropy is a necessary condition for physical consistency of the low-temperature solution. It then means that the low-temperature equilibrium state for $p<\infty$ must contain the Parisi continuous order-parameter function $x(\mu)$ with $\beta > \mu_{1}> \mu_{0}$. It can also be seen from the asymptotic free energy for $p\to\infty$ that reads
\begin{multline}\label{eq:FE-infty}
f^{(p\to\infty)}_{T}(q,\chi_{1},\mu_{1}) =  - \frac{1}{4T} \left[ 1 - \left(q + \chi_{1}\right)( 1 - \ln \left(q + \chi_{1}\right)\right] - \frac 1{\mu_{1}}\ln \left[2\cosh(\mu_{1}h)\right]  \\  - \frac{\mu_{1}}{4} \left[ \chi_{1} - \left(q + \chi_{1}\right) \ln \left(q + \chi_{1}\right)\right]   -  \frac{\mu_{1}q}{4}\left[ \ln q  + p \left(1 - \tanh^{2}(\mu_{1}h)\right) \right]  \ ,
 \end{multline}
giving the leading-order solution for the variational parameters $\chi_{1},\mu_{1}$ and $q$.  The first two parameters are of order one while the latter is exponentially small for large $p$,  
\begin{subequations}\label{eq:stationary-asymptotic}
\begin{align}
\chi_{1} &=1 - q \ ,\\ 
q &= \exp\{- p(1 - \tanh^{2}(\mu_{1}h) \}\ , \\
 \mu_{1} &= 2 \sqrt{\ln \left[2\cosh(\mu_{1}h)\right]  - h\tanh(\mu_{1}h)} \ .
\end{align}
\end{subequations}
The above nontrivial solution holds only if $\beta > 2 \sqrt{\ln \left[2\cosh(\beta h)\right]  - h\tanh(\beta h)}$ (low-temp\-erature phase), otherwise $\mu_{1}= \beta$  and $\chi_{1} + q=0$ (high-temperature phase). To derive an equation for the order-parameter function $x(\mu)$ one needs to include the next-to-leading order contributions. To go beyond the leading asymptotic order one can use the Landau-type theory for the order-parameter function developed in Ref.~\cite{Janis:2013aa}. Note that the asymptotic solution of Eqs.~\eqref{eq:stationary-asymptotic} with $x(\mu) = 0$ suffers from a negative entropy as can be seen from Eq.~\eqref{eq:S-Tlow} and is plotted in Fig.~\ref{fig:1RSB-entropy}.  Note that the transition to the ordered phase in the $p$-spin glass is discontinuous and hence, an asymptotic expansion below the transition temperature is not applicable. Only an asymptotic expansion $p\to \infty$ makes sense. 

\section{Conclusions}

The mean-field models of spin glasses have equilibrium states demanding nonstandard tools to describe them quantitatively. Moreover, the replica trick standardly used is suspicious and it is difficult to understand and give a physical meaning to the order parameters and the functions with the replica variables in this construction.  The standard way to reach the equilibrium states is to use the replica trick, derive representations with the replica variables and then try to give the results a physical meaning. The reason for using the replicas remains nevertheless veiled in this way and the physical meaning of the replica-symmetry breaking unclear \cite{Parisi:1983aa,Mezard:1984aa}.  One has to analyze the physical reasons for instabilities and negative entropy in non-replicated mean-field solutions to find the genuine reason why the true equilibrium of the mean-field spin-glass models cannot be reached without replicas.      

The most prominent feature of the mean-field spin-glass models is that the combination of randomness and frustration of the spin exchange leads to \textit{ergodicity breaking} that is not accompanied by any symmetry breaking in the generic Hamiltonian. To restore ergodicity in the glassy phase  one can use replications of the phase space of the dynamical variables. A small inter-replica interaction  is introduced as a perturbation and the linear response of the system is calculated. If the linear response (replica independence or symmetry) is broken, replica variables influence the behavior of the original system, and inter-replica interactions become non-zero. The replication is then used hierarchically till one restores at least local thermodynamic homogeneity. The principal step in this procedure is to select an adequate symmetry-breaking of the replicated variables so that to make thermodynamic potentials analytic functions of the originally integer replication index. Only then it is possible to test and restore thermodynamic homogeneity and ergodicity. If infinitely-many replications are needed the Parisi solution with the continuous replica-symmetry breaking is obtained. Real replicas allow us to restore ergodicity in hierarchical steps by breaking successively independence of the replicated spaces. 

The solution of the full mean-field models of spin glasses in unreachable. One has to resort to approximations or perturbative expansions.  We applied the real replicas on the mean-field Ising, $p$-state Potts and $p$-spin glass models and calculated their asymptotic solutions below the transition temperature to the glassy phase.  We thereby demonstrated that the Ising spin glass below the transition point is described by a solution with infinitely-many replica hierarchies with the continuous order-parameter function of Parisi. While the Ising spin glass is known to have continuously broken replica symmetry in the equilibrium state, the Potts and $p$-spin glasses allow for locally stable solutions with a one-level discrete replica-symmetry breaking. Since the solution with the continuous replica-symmetry breaking exists independently of stability of the solutions with finite-many replica hierarchies,  the continuous RSB and 1RSB coexist in the $p$-state Potts and the $p$-spin glass models. In both cases for $p<\infty$ the 1RSB state leads to negative entropy at very low temperatures and the ultimate equilibrium state for the mean-field spin-glass models  breaks the replica symmetry in a continuos form as suggested by Parisi in the Ising model. Our analysis indicates that a continuous RSB is indispensable to keep entropy  non-negative down to zero temperature.  Spin reflection symmetry is hence not substantial for the existence of a solution with a continuous replica-symmetry breaking as was previously assumed.

\section*{Acknowledgement}

I would like to thank my collaborators A. Kl\'\i\v c and A. Kauch for their numerical calculations and the figures I used  in these lecture notes. I also thank the Fulbright Commission for financing my stay at Louisiana State University in Baton Rouge where I compiled this contribution.


\clearpage

\clearchapter


\end{document}